# Generating tightly focused perfect optical vortex for ultra-secure optical encryption

*Qingshuai Yang, Zijian Xie, Mengrui Zhang, Xu Ouyang, Yi Xu, Yaoyu Cao, Sicong Wang, Linwei Zhu and Xiangping Li\**

Q. Yang, Z. Xie, M. Zhang, Dr. X. Ouyang, Prof. Y. Cao, Prof. S. Wang, Prof. X. Li, Guangdong Provincial Key Laboratory of Optical Fiber Sensing and Communications, Institute of Photonics Technology, Jinan University, Guangzhou 510632, China.
Email : xiangpingli@jnu.edu.cn
Prof. Y. Xu, Advanced Institute of Photonics Technology, School of Information Engineering, and Guangdong Provincial Key Laboratory of Information Photonics Technology, Guangdong University of Technology, Guangzhou 51006, China
Prof. L. Zhu, School of Physics and Optoelectronic Engineering, Ludong University, Yantai 264025, China

Light's orbital angular momentum (OAM) with inherent mode orthogonality has been suggested as a new way to the optical encryption. However, the dependence of annular intensity profiles on the topological charge complicates nanoscale light-matter interactions and hampers the ultra-secure encryption application. In this paper, we demonstrate ultra-secure image encryption by tightly focusing perfect optical vortex (POV) beams with controllable annular intensity profiles and OAM states. A simple scheme composed of single spatial light modulator is proposed to generate radius-controllable POV in tightly focused beams. Such focused POV beams with identical intensity profiles but varied OAM states are applied to disorder-coupled gold nanorod aggregates to selectively excite electromagnetic hot spots for encoding information through photothermal deformation. As such, ultra-secure image encryption in OAM states of POV beams in combination with different polarizations can be achieved. Our results lay the ground for diverse nanophotonic applications harnessing the OAM division of POV beams.

## 1. Introduction

Recently, optical vortices have attracted significant research interests owing to the well-defined on-axis orbital angular momentum (OAM) they can carry.[1] A helical wavefront of $\exp(i\ell\phi)$ with multiple of $2\pi$ phase accumulation winding around the beam center resembles its characteristic properties.[2] This prominent feature has impinged diver photonic applications including optical trapping,[3-5] superresolution imaging,[6] lasing,[7-9] and optical communications.[10, 11] In particular, the inherent mode orthogonality of OAM states has been suggested as an excellent information carrier for multiplexing to boost the information capacity of both optical communications,[12, 13] and holographic encryption.[14-16] However, the annular intensity profile and peak intensity vary as a function of topological charges of vortex beams. The interdependence complicates nanoscale light-matter interactions with varied both local intensity and OAM state and this property becomes problematic in applications that require to couple multiple OAM beams into fixed spatial modal distributions.

In this regard, the concept of perfect optical vortex (POV) whose annular intensity profiles of the generated beam are immune to the variation of topological charge has been introduced.[17] An idea POV beam can be treated as the Fourier transform of a Bessel beam. The general approaches to create such POV beams generally involve the superposition of an axicon phase function with a vortex phase, which has been implemented through meta-surface elements,[18] digital micromirror device (DMD),[19] conical axicon,[20] and liquid-crystal spatial light modulators (SLMs).[21, 22] Unfortunately, these methods are only demonstrated effective in paraxial conditions. The presence of an axicon phase leads to defocusing effect in the propagation axis and the shifted focal spot is often understated with degraded both intensity profiles and ring radius, especially in applications where tightly focused POV beams are on-demand.

In this paper, we demonstrate POV with controlled annular radius and peak intensities in tightly focused conditions. The complex field of Fourier transfer of the focused POV expressed as a diffraction-limited annular ring intensity profile superposed with a vortex phase is implemented through a single SLM approach. It allows the generated POV beam with controlled annular radius and arbitrary OAM states in the focal plane. As a proof-of-principle, such tightly focused POV beams are applied to gold nanorod aggregates for ultra-secure image encryption in both topological charges and polarizations through a selective photothermal deformation process.

## 2. Results and discussion

### 2.1. Theoretical and experimental verification of tightly focused POV

The complex amplitude expression of an ideal POV with topological charge $l$ is given by:[17]

$$E(\rho, \theta) = \delta(\rho - \rho_0)\exp(il\theta) \quad (1)$$

where $(\rho, \theta)$ are the polar coordinates in the beam cross section, $\delta(x)$ is Dirac $\delta$-function, and $\rho_0$ is the radius of POV. The POV with this complex amplitude distribution can be obtained using the Fourier transform property of an ideal Bessel mode, which is given by:[23]

$$B(r, \varphi) = J_l(\alpha r)\exp(il\varphi) \quad (2)$$

where $\alpha$ is the wave vector in the transverse direction, $J_l(\alpha r)$ is an $l$th order Bessel equation of the first kind. In experiment, if we want to generate a POV by means of Eq. (2), it must be bounded by a circular aperture of radius R:[24]

$$B_1(r, \varphi) = circ(\tfrac{r}{R})J_l(\alpha r)\exp(il\varphi) \quad (3)$$

where

$$circ\left(\frac{r}{R}\right) = \begin{cases} 1, & r < R, \\ 0, & r > R. \end{cases}$$

For the field distribution in Eq. (3), its Fourier transform in polar coordinates is given by:[24]

$$U(\rho, \theta) = (-i)^{l+1}\left(\frac{k}{f}\right)e^{il\theta}\int_0^R J_l(\alpha r)J_l\left(\frac{kr\rho}{f}\right)rdr$$

$$= (-i)^{l+1}\left(\frac{kR}{f}\right)e^{il\theta} \times \left[\frac{\alpha J_{l+1}(\alpha R)J_l(XR) - XJ_l(\alpha R)J_{l+1}(XR)}{\alpha^2 - X^2}\right] \quad (4)$$

where $X = k\rho/f$, $k = 2\pi/\lambda$ is the wavenumber, $\lambda$ is the wavelength of the incident beam, $f$ is the focal length of an ideal Fourier transform lens. Using the orthogonality of Bessel functions to simplify Eq. (4), it can be seen that the maximum radius of the focus ring is the same as Eq. (1), being independent of the topological charge. If we want to apply it to a high NA focusing system, we must consider the Fourier transform (FT) effect of the high-NA objective of the weighted field: [25]

$$E(x, y, z) = FT\{G(\Theta, \phi)\}$$
$$= FT\{P(\Theta)E_t(\Theta, \phi) \exp(jkz\cos\Theta)/\cos\Theta\} \qquad (5)$$

where $FT\{\cdot\}$ represent the Fourier transform; $P(\Theta)$ is the apodization function; $\Theta$ is the converge angle, which is related to the aperture radius, the numerical aperture and refractive index of the immersion medium; $E_t(\Theta, \phi)$ is the transmitted field. Then, when the incident optical beam is modulated into the field of Eq. (5), the electric field distribution in the focal plane can be written as:

$$FT\{U(x_0, y_0) \times G(\Theta, \phi)\} \propto \exp\left(-\frac{(\rho-\rho_0)^2}{\Delta\rho^2}\right)\exp(il\psi) \qquad (6)$$

where $\rho_0$ and $\Delta\rho$ denote, respectively, the radius and width of the annulus, and $\psi$ is the azimuthal angle in the focal plane of the objective.

The characteristic of Eq. 6 is that the topological charge will not affect the amplitude distribution, which satisfies the definition of POV. Therefore, we can generate a POV by modulating the optical field according to Eq. 4. There are both amplitude modulation and phase modulation in Eq. 4, which can be expressed as $U(x_0, y_0) = A(x_0, y_0)\exp(i\phi(x_0, y_0))$ in the cartesian coordinate. The amplitude and phase distributions of the complex field are shown in the **Figure 1**a and Figure 1b. In this case, the intensity radius $R = 3\ \mu m$, and the topological charge $l = 5$.

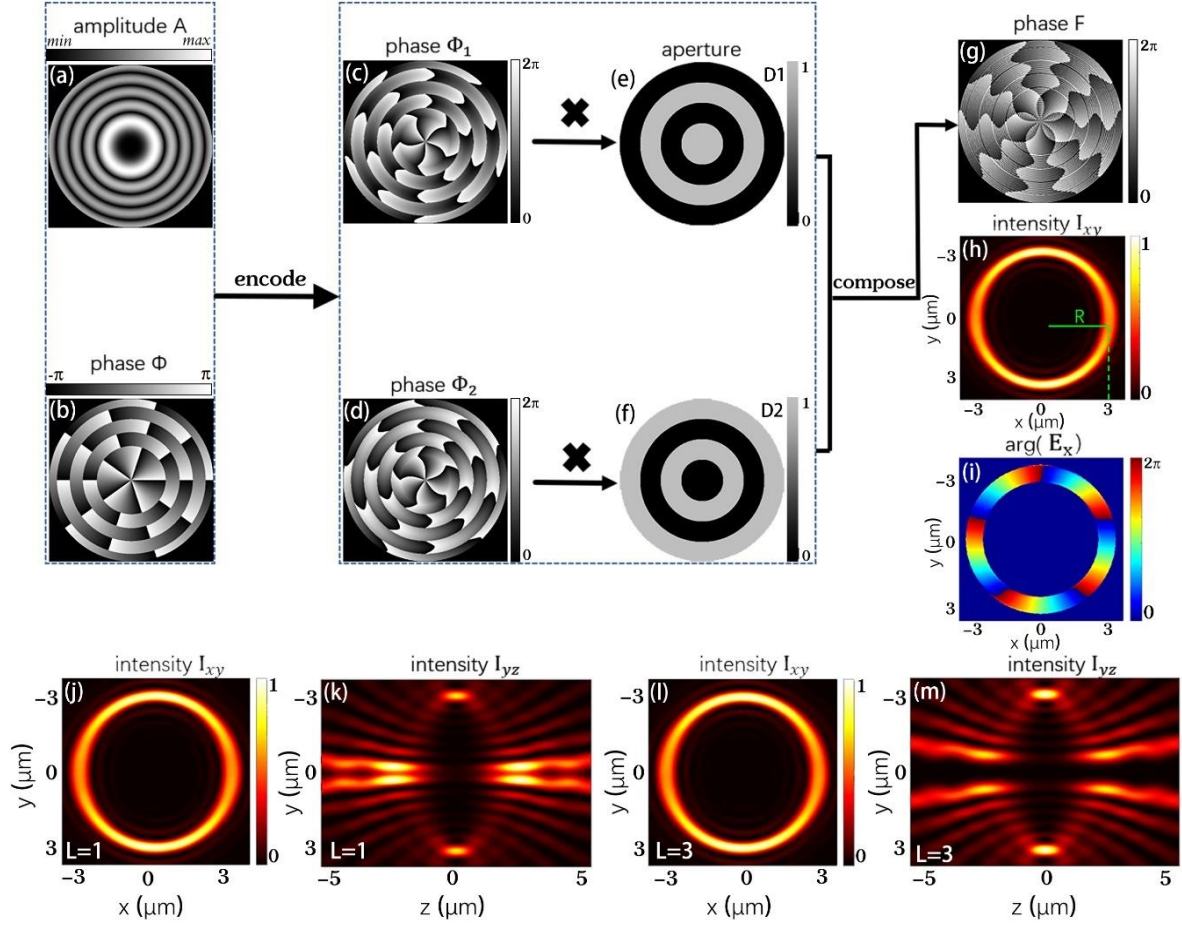

**Figure 1.** a) Amplitude and b) phase distributions of the complex field in Eq. 4. c) - d) Two phase-only distributions correspond to Eq.7. e) - f) a pair of complementary ring-shaped apertures. g) the final synthesized phase distribution. h) - i) show the intensity distribution in the $xy$ cross-section and the phase whose topological charge $l = 5$. j) - m) are the intensity distributions in the $xy$ and $yz$ cross-sections, and the corresponding topological charges $l = 1$ and $l = 3$. Simulated parameters of tightly focused system: x polarization, λ=800 nm, NA = 0.95, radius R =3 μm.

In the experiment, we try to use a phase-only SLM to generate complex amplitude modulations, in which amplitude and phase are modulated simultaneously. Firstly, we rewrite the complex field expression as the superposition of two phase-only modulations: $U(x_0, y_0) = A_{max}/2 \cdot \exp(i\phi_1(x_0, y_0)) + A_{max}/2 \cdot \exp(i\phi_2(x_0, y_0))$, where the specific expressions are:

$$\begin{cases} \phi_1(x_0, y_0) = \phi(x_0, y_0) + \arccos[A(x_0, y_0)/A_{max}] \\ \phi_2(x_0, y_0) = \phi(x_0, y_0) - \arccos[A(x_0, y_0)/A_{max}] \end{cases} \quad (7)$$

The corresponding calculation results of two phase-only distributions are displayed in Figure 1c – Figure 1d. Then, in order to use a phase pattern to generate the complex field, the two kinds of different phase information must be superimposed on a single pattern. Here, we designed a pair of complementary ring-shaped apertures, $D_1(x_0, y_0)D_2(x_0, y_0)$, to combine the two different phase distributions. As shown in Figure 1e – Figure 1f, they are binary structures with alternating amplitude of 0 and 1, in which each ring has the same width. Superimpose the previously obtained phase distribution through complementary ring-shaped apertures:

$$F(x_0, y_0) = \phi_1(x_0, y_0)D_1(x_0, y_0) + \phi_2(x_0, y_0)D_2(x_0, y_0) \tag{8}$$

The superimposed phase is displayed in Figure 1g. Based on the synthetic phase, we can realize the complex optical field. Finally, a POV is created in the tightly focus system.

Figure 1h shows the intensity distribution in the focus plane of a high NA objective lens when the incident optical field is modulated by phase $F$. It can be seen that the intensity profile of the optical field is a ring shape, and the radius is consistent with the theoretically designed value, indicating that the radius is controllable by this method. Simultaneously, its phase distribution of $E_x$ exhibits 5 fold of $2\pi$ phase accumulation winding around the beam center, coinciding with theoretically designed topological charge $l = 5$ (see Figure 1i). Figure 1j and Figure 1l show the intensity distributions of POV with the same R on the $xy$ cross-section, whose topological charge are $l = 1$ and $l = 3$, respectively. Their intensity distributions on the $yz$ cross-section are shown in Figure 1k and Figure 1m. We can see that dominant intensity is located at the focal plane (Z = 0), and this part of intensity increases as the topological charge increases.

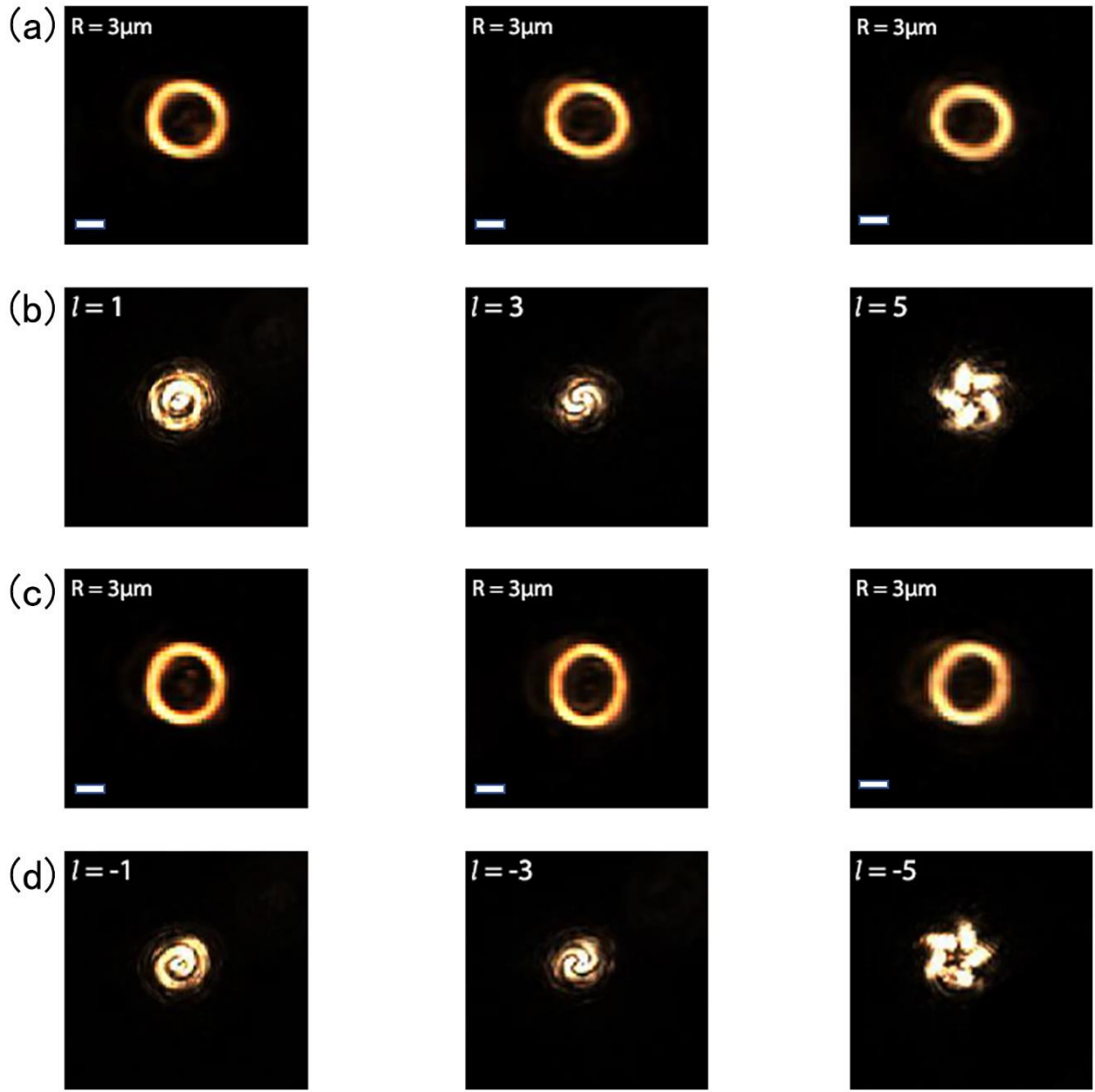

**Figure 2.** a) and c) are intensity patterns of perfect optical vortex with different topological charges on the focal plane (Z = 0). b) and d) are the corresponding coaxial interference patterns. Scale bar: 3μm

To experimentally verify the tightly focused POV, we conducted an interference experiment with our homemade setup. **Figure 2**a and Figure2c show the POVs generated on the focal plane of the objective lens, with various topological charges ($l = \pm1, \pm3, \pm5$) and the same radius (R = 3 μm). Then we manage to let another unmodulated Gaussian beam interfere with the POV to confirm the vortex nature of the ring beam.[21] Figure 2b and Figure2 d show the result of the interference experiment, and the spiral fringe pattern confirms the presence of OAM in the POV. The number of fringes represents the topological charge $l$, and its direction of rotation decides the sign. These results solidly confirm that the tightly focused POV can have the same radius while with controllable OAM states. Furthermore, we show that this method can generate POVs with different radii while keeping the topological charge constant (Supplementary Figure S1).

**2.2 Simulation and experimental results of optical storage**

When these tightly focused POV interact with gold nanorod aggregates, the focal polarization distributions play an important role. It has been shown that the vortex beam can introduce OAM-dependent spatial rotations of the focal polarization ellipses and hence synthetic helical dichroism in individual gold nanorods.[26] Based on the vectorial diffraction integral theory, we can calculate and analyze the electric field components, $Ex\ Ey\ Ez$, of the focal electric field, and further express the polarization state of each point through the polarization ellipse. The polarization ellipse of such focused POV in the $xy$ cross section can be expressed as:[27]

$$\frac{E_x^2}{E_{0x}^2} + \frac{E_y^2}{E_{0y}^2} - 2\frac{E_x}{E_{0x}}\frac{E_y}{E_{0y}}\cos(\psi) = \sin(\psi) \tag{9}$$

where $E_x = E_{0x}e^{-i\omega t+ikz+i\delta_x}$ and $E_y = E_{0y}e^{-i\omega t+ikz+i\delta_y}$ are the complex amplitude expression of electric field components; $\psi = \delta_x - \delta_y$ is the phase retardation between $E_x$ and $E_y$. Following Eq. (9), the ellipse orientation angle and eccentricity at each point within the cross-section plane can be calculated. Then we superimpose and display the polarization ellipse and intensity distribution corresponding to each point on the defocused plane (z = 100 nm), as shown in the first row of **Figure 3**a. The calculation parameter is the same as the previous setting, and we select twenty-four columns of data at equal intervals to calculate the corresponding polarization ellipse. Similarly, we can calculate the polarization states distribution in the $yz$ and $xz$ cross section planes, which are illustrated in Figure 3a. It is seen that the orientations of polarization ellipses manifest a spatial dependence. In addition, the rotation of the polarization ellipse is dispersed at the same position of the focal region which is determined by the topological charge of the tightly focused POV (Supplementary Fig. S2).

Consequently, the POVs with controllable radius and different OAM states can be applied to gold nanorod aggregates to excite localized electromagnetic hotspots to encrypt different information.

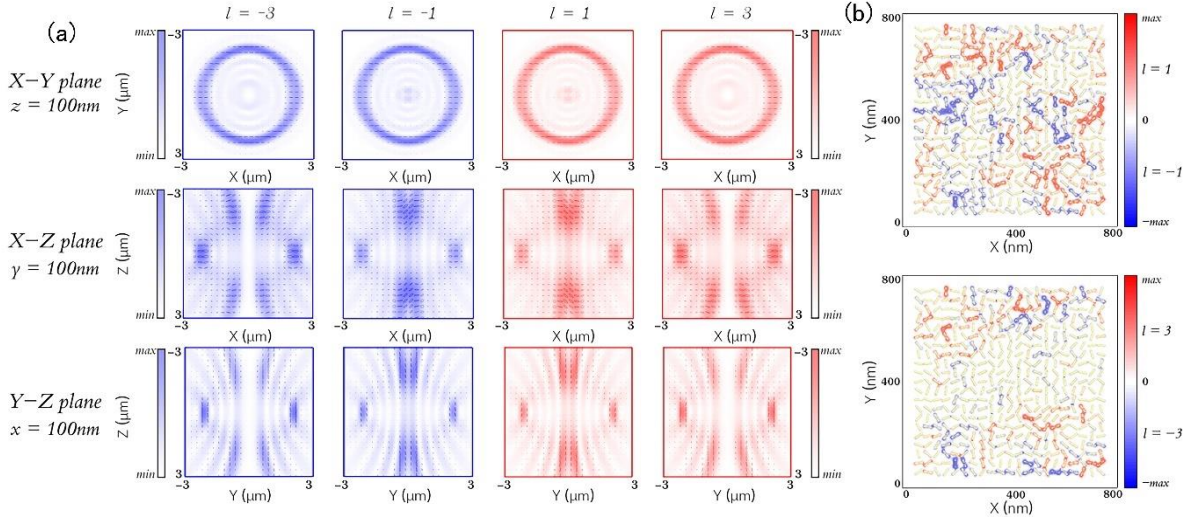

**Figure 3.** a) The polarization ellipse and intensity distributions of the $xy$、$xz$、$yz$ cross sections in the focal region. The red and blue ellipses represent the polarization state of the points in the focal regions when l = -3, -1, 1 and 3, respectively. b) The red and blue color denote random hot spots distribution when disordered gold nanorods interact with tightly focused POV beams with different topological charges ($l = \pm 1$, $l = \pm 3$), which are simulated by FDTD.

In order to provide an intuitive illustration, we used a simplified model to qualitatively evaluate the response of a disordered gold nanorod assembly under the excitation of a tightly focused POV.[26, 28] The simplified model consists of 441 gold nanorods, which are 34 nm in length and 8 nm in diameter. The gold nanorod array contains $21 \times 21$ randomly oriented gold nanorods whose centers form a square array with a lattice constant of 34 nm. The electric field distribution of gold nanorods can be calculated by the finite-difference time-domain method (FDTD) and the non-uniform grid with the maximum mesh step of 2 nm as well as perfectly matched layer (PML) boundary conditions are used in the numerical simulation. By calculating the intensity distribution generated by the interaction between the source which has been imported into the POV's electric field distribution and the disordered gold nanorods model, we can intuitively see the excitation of localized electromagnetic hot spots by POV with variant topological charges, as shown in Figure 3b. It is clearly seen that the recording beam with different topological charges can excite different hot spots constituted by different gold nanorods denoted by red and blue color. When raising the beam power, the gold nanorods nearby these hot spots can be selectively photothermally reshaped to provide a mechanism for optical encryption in the OAM states of the POV.

To experimentally demonstrate ultra-secure optical image encryption by tightly focused POV, we prepared samples of gold nanorod aggregates. Gold nanorods with an initial optical density of 90 were mixed with 10 wt.% Poly(vinyl alcohol) polymer and self-dried at room temperature. In **Figure 4**a, we show eight images encrypted at the same spatial region by the eight types of combination with polarization and topological charge, which have been marked in the corresponding positions. Each image consists of 30 by 30 pixels with a pixel separation of 1.5 $\mu m$. The images were encrypted through photothermal reshaping of gold nanorods in a homebuilt optical setup (Supplementary Fig. S3). The exposure time for each pixel was 20 $ms$ and the laser power used for encryption was optimized to minimize the cross-talk. The recorded patterns were retrieved through the contrast in nonlinear upconversion luminescence[29, 30] between encrypted and non-encrypted regions. On the basis of normalizing the fluorescence intensity, we select suitable thresholds according to the bit error rate (Supplementary Fig. S4) to binarize the image, as shown in Figure 4a. Figure 4b presents the fidelities of retrieved images which are calculated based on the raw nonlinear upconversion luminescence data. It can be seen that when retrieving the encrypted images by the POV with identical OAM states and polarizations to the encrypting beams the fidelity can be far exceeding 99%.

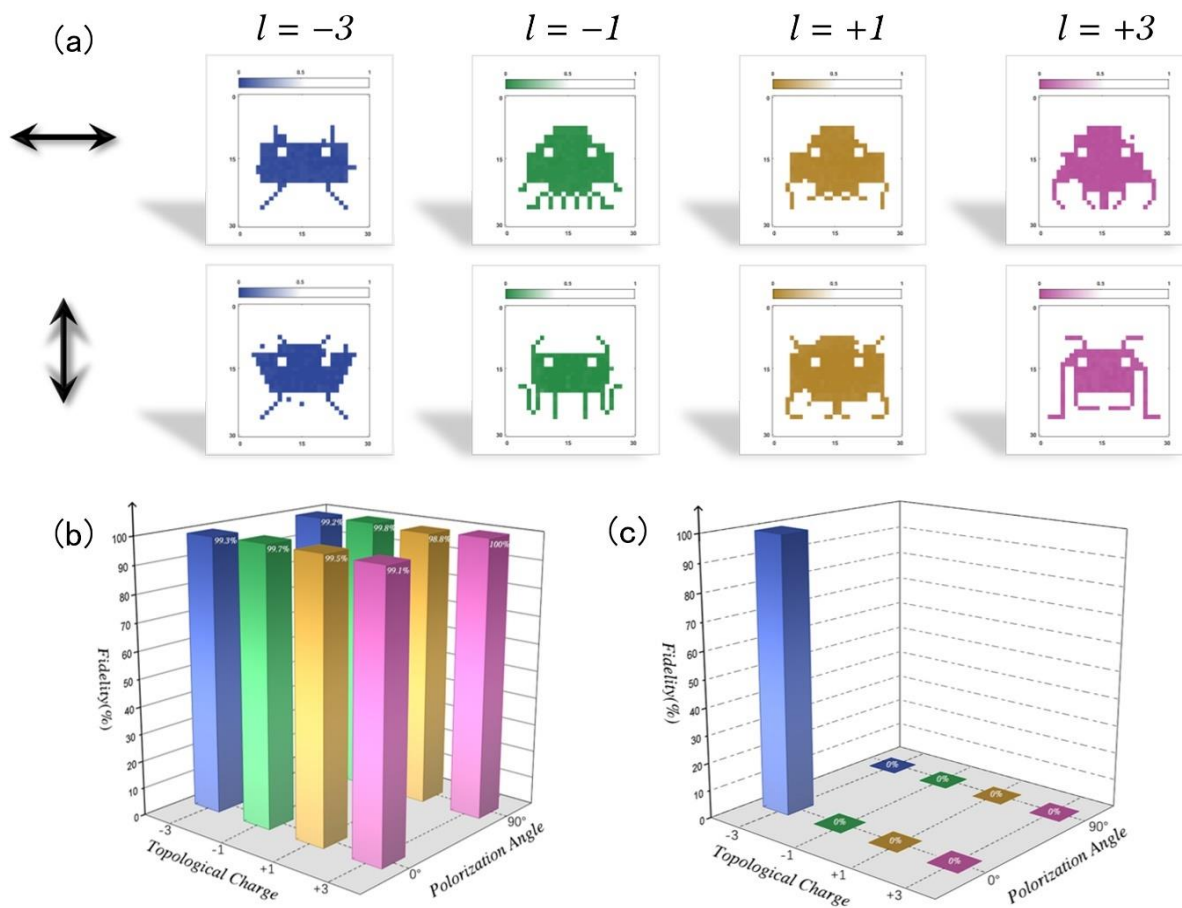

**Figure 4.** a) Schematic of the optical image encryption in a spatial region ($45 \times 45$ $\mu m^2$ with $30 \times 30$ pixels). The eight images are the retrieved results using eight different polarization and topological charge combinations, which are outlined in b). The fidelity of these eight optically encrypted images in a). c) Analysis of fidelity of retrieved images using different combination of polarizations and topological charges.

Furthermore, we used a POV beam with topological charge $l = 1$ and polarization $P = 0°$ to record an image, and then point-by-point scan the nonlinear upconversion luminescence signal of the recording area with a POV beam of different combinations of polarizations and topological charges. The corresponding fidelity is shown in Figure 4c. It is clearly seen that the encrypted information can only be retrieved when read by the POV with the same parameter that was employed during the encryption, otherwise the information is read out as noise with fidelity approaching null (Supplementary Fig. S5). These results indicate that ultra-secure optical image encryption by the generated POV beams with controllable OAM states and polarization is feasible.

## 3. Conclusion

In summary, we propose a novel method to generate POV in the focal plane of tightly focused systems and we can control the radius and the topological charge based on a phase-only formula. Based on the complex expression of the Fourier transform of the ideal Bessel beam, we use a single phase-only SLM to achieve a complex field and focus this field onto the focal plane where the created POV is used to encryption. The image encryption has been demonstrated through the interaction between focused POV and disorder-coupled gold nanorod. Additionally, this novel approach can be used to generate other types of POV with different shape of the intensity distribution, such as elliptic perfect optical vortex,[31, 32] fractional perfect optical vortex.[33] In theory, the physical dimension of orbital angular momentum has boundless orthogonal states. It is anticipated that the combination of orbital angular momentum and other physical dimensions can further increase the storage capacity of optical memory technology. We envision its huge potentials of application of this degree of freedom in different fields, such as multiplexed data storage, optical communication, and quantum entanglement, etc.

**Supporting Information**
Supporting Information is available from the Wiley Online Library or from the author.

**Acknowledgements**

The authors would like to acknowledge the financial support from the National Nature and Science Foundation of China (Grant Nos. 61522504 and 91750110), the Guangdong Provincial Innovation and Entrepreneurship Project (Grant 2016ZT06D081 and 2019ZT08X340), the Research and Development Plan in Key Areas of Guangdong Province (2018B010114002), the Pearl River Nova Program of Guangzhou (No. 201806010040), National Natural Science Foundation of China (NSFC)(61975066), National Natural Science Foundation of China (NSFC)(61875073), National Natural Science Foundation of China (NSFC)(91750110) and National Natural Science Foundation of China (NSFC)(62174073).

Received:
Revised:
Published online:

**Conflicts of interest**
The authors declare no conflicts to interest.

Supporting Information

**Generating tightly focused perfect optical vortex for ultra-secure optical encryption**

*Qingshuai Yang, Zijian Xie, Mengrui Zhang, Xu Ouyang, Yi Xu, Yaoyu Cao, Sicong Wang, Linwei Zhu, and Xiangping Li\**

**1. Intensity patterns of the generated perfect optical vortex beams with different radius**

In **Figure S1**, we present POV with different radius which could be realized by modifying the parameters of the phase-only formula. The corresponding coaxial interference patterns obtained by making the POV interfere with Gaussian beam. The number of spiral fringes is the same as the topological charge, and the shapes will change with the radius.

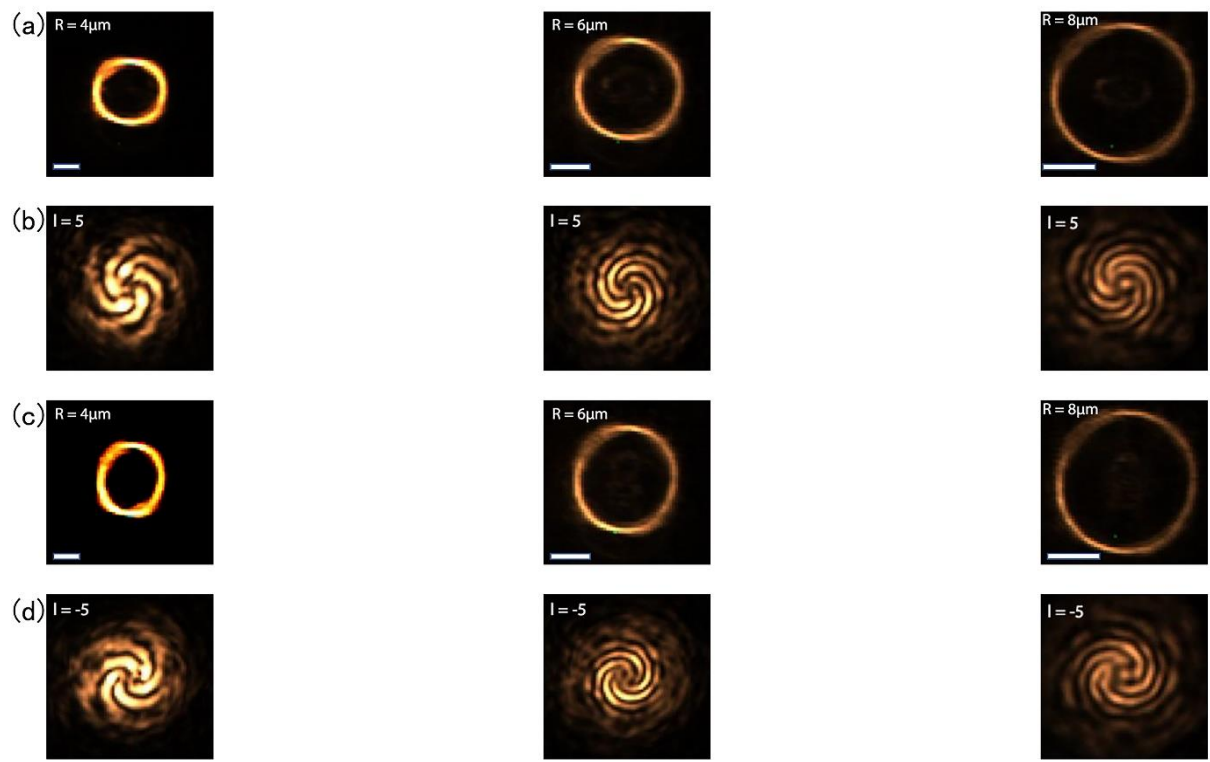

**Figure S1.** a) and c) are the POV's intensity distributions in the focal plane with various radii (R = 4μm, 6μm, 8μm), corresponding to $l=5$ and $l=-5$, respectively. b) and d) are interference patterns of perfect vortex beam and Gaussian beam. Scale bars in three columns denote: 4 μm, 6 μm, 8 μm.

**2. The inclination of polarization ellipses on the $xy$、$xz$、$yz$ cross sections**

In **Figure S2**a, we show a schematic diagram of the inclination of the polarization ellipse, which represents the angle ($\theta$) between the major axis of the ellipse and the horizontal axis. The expression of the inclination angle ($\theta$) can be defined as:

$$tan2\theta = \frac{2E_{0x}E_{0y}cos\psi}{E_{0x}^2 - E_{0y}^2} \quad (1)$$

Where $\psi$ is phase difference between $E_x$ and $E_y$; $E_{0x}$ and $E_{0y}$ are the amplitudes of $E_x$ and $E_y$, respectively.

We have shown the polarization ellipse distributions in the corresponding sections in Figure 3a. To highlight the difference of electric fields with topological charge of different symbols and values, we plot the inclination of polarization ellipses corresponding to different sections, as shown in Figure S2b. It is clearly seen that there are obvious changes between the cross sections in the electric field of different topological charges and the calculation result has a variation range of 180 degrees, which is consistent with the definition.

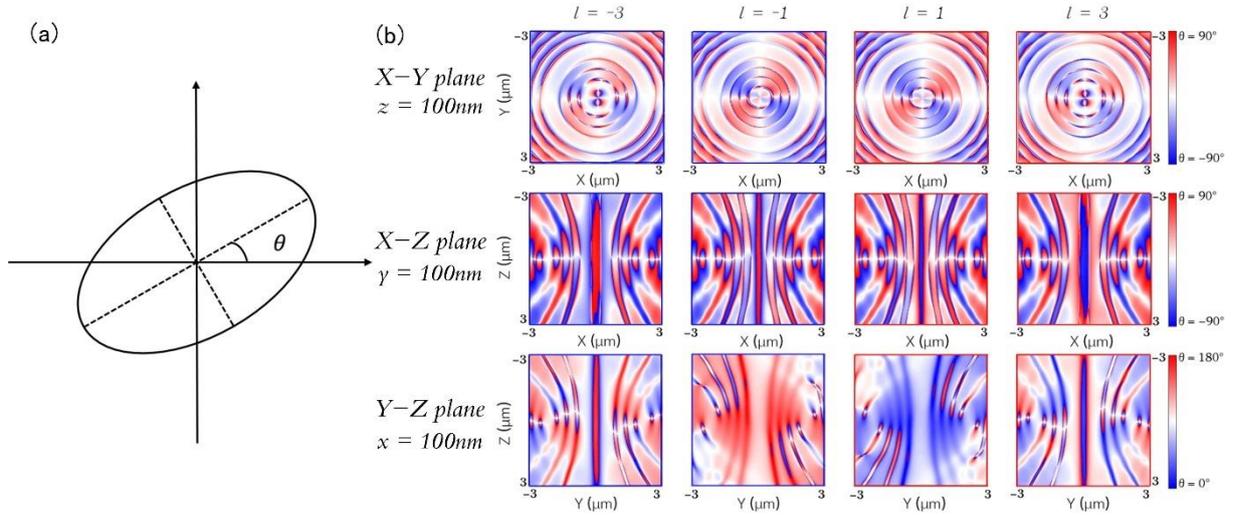

**Figure S2.** a) Schematic diagram of the ellipse inclination $\theta$. b) The distribution of inclinations $\theta$ on the $xy$、$xz$、$yz$ cross-sections, which are offset by 100nm from the focal point of the light field.

## 3. The optical setup

We perform the optical encryption experiment with our homemade setup shown in **Figure S3**. The 800nm fs laser from a Ti:sapphire oscillator (Chameleon, Coherent) is expanded and collimated onto a half-wave plate and a Glan prism to adjust the power of laser and control the direction of polarization of the outgoing light. The output light oblique incident into the phase-only space light modulator (HOLOEYE, Pluto LC-R2500) and then through the 4F system and dichromatic mirror (FF735-Di02-25 × 36, Semrock). Finally, the laser focused to the sample placed on a three-dimensional stage (P-563.3CD, Physik Instrumente) by an objective. The nonlinear upconversion luminescence of the sample was collected by the same objective and detected by an avalanche photo-diode (SPCM-AQRH-14-FC, Excelitas Technologies), which is placed in the position of the CCD in the diagram.

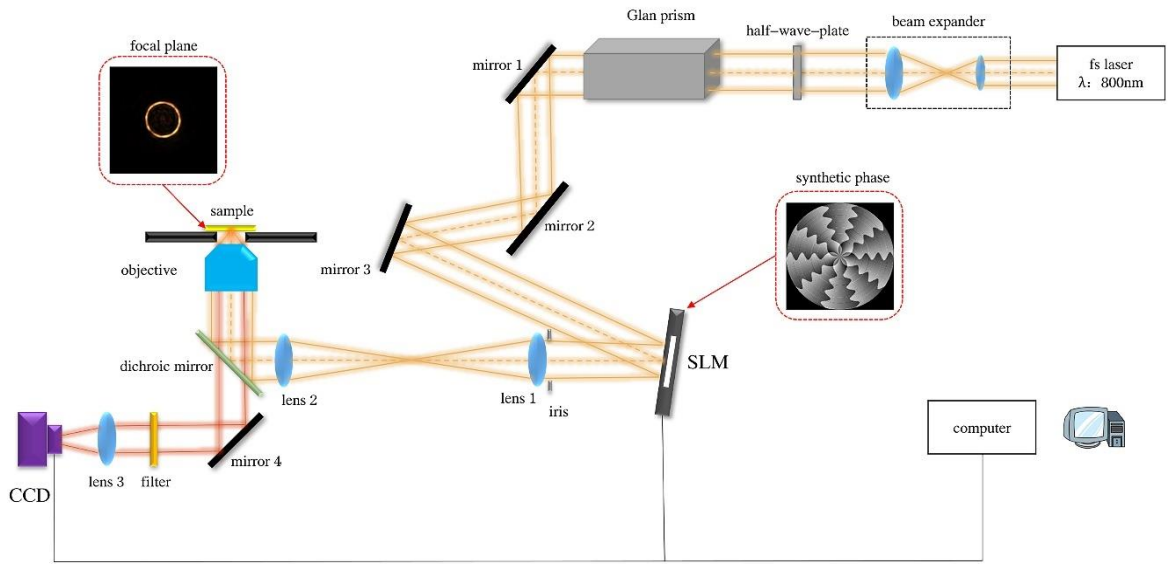

**Figure S3.** The schematic of the optical setup used in the optical encryption experiment. fs laser: femtosecond laser; SLM: spatial light modulator; CCD: charge coupled device.

### 4. The data processing method

We take one of the pictures of image encryption experiments as an example to illustrate the process of original data processing, which is shown in **Figure S4**. Firstly, we code a program to scan the recording area to acquire the corresponding nonlinear upconversion luminescence intensity distribution, as shown in Figure S4a. Then, we calculate the bit error rate corresponding to different thresholds, and select the appropriate threshold to binarize the nonlinear upconversion luminescence image, which is shown in Figure S4b. The bit error rate (BER) can be defined as:

$$e = \frac{\sum_m \sum_n |A_{mn} - B_{mn}|}{N} \qquad (2)$$

where $|A_{mn} - B_{mn}|$ denotes the error bit between the original data and the extracted binarization data of all the information units. N is the total number of information units.

For the statistical graph of the nonlinear upconversion luminescence intensity distribution of each pixel in Figure S4a, there are two Gaussian distributions corresponding to the recorded points and the non-recorded points, as shown in Figure S4d. Figure S4b shows the binarized pattern at the threshold intensity Ith = 0.2894. The bit error rate corresponding to this set of data is 0.0044 and this result demonstrate that our concept can be used for encryption with high-quality and low crosstalk.

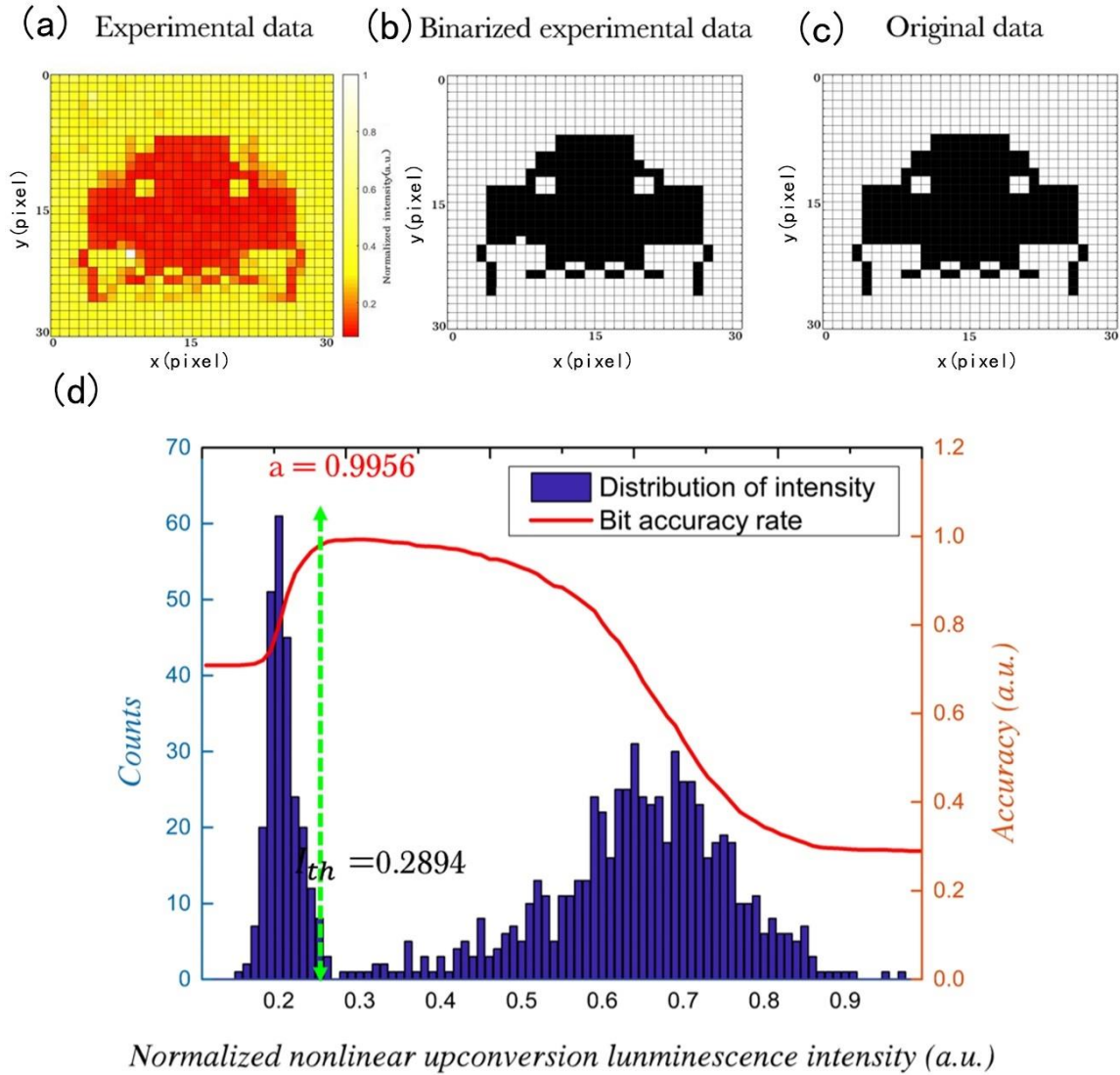

**Figure S4.** a) The pattern obtained by detecting the nonlinear upconversion luminescenceintensities of all the information units. The size of the recording region was 45 × 45 μm with 30 ×30 pixels. b) The result of binarizing the nonlinear upconversion luminescence image by selecting an appropriate threshold. c) The original pattern used for data recording. d)

The distribution of the nonlinear upconversion luminescence intensities of all the information units in the recording region. The calculated fidelity (a = 0.9956) and threshold (Ith = 0.2894) for the extracted pattern is also provided.

## 5. The nonlinear upconversion luminescence intensity distributions

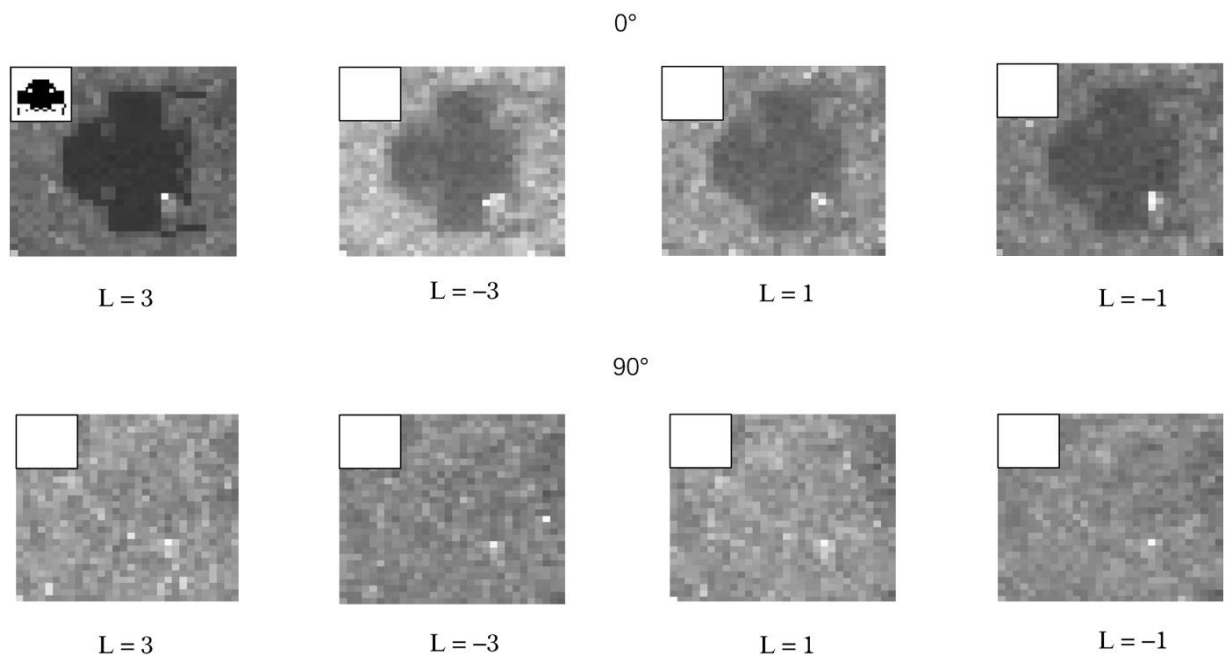

**Figure S5.** The nonlinear upconversion luminescence images corresponding to every polarization and topological charge combinations in Figure 4c. The insets represent the reconstructed image after processing.